\documentclass[%
reprint,
amsmath,amssymb,
pra,
]{revtex4-2}

\usepackage{graphicx}
\usepackage{dcolumn}
\usepackage{bm}
\usepackage{braket} 
\usepackage{tikz}
\usepackage{amsmath}
\usepackage{amsthm}

\begin{document}

\newcommand{\vecg}{\boldsymbol}
\renewcommand{\vec}{\textbf}

\title{Two-qutrit Werner state is always local}

\author{Pawe\l{} Caban and Muhammad Mohsin}
\affiliation{%
Department of Theoretical Physics, University of \L\'od\'z \\
Pomorska 149/154, PL-90-236 \L\'od\'z, Poland
}%
\email{pawel.caban@uni.lodz.pl,\\ muhammad.mohsin@edu.uni.lodz.pl}

\date{\today}

\begin{abstract}
In our paper, we show that a two-qutrit Werner state admits 
a local hidden-variable model for all values of the parameter 
in a scenario involving two projective measurements per site, 
with three outcomes per measurement. 
To this end, we derive a simple sufficient condition under which 
a two-qutrit state satisfies the Collins–Gisin–Linden–Massar–Popescu
inequality for arbitrary choices of Alice’s 
and Bob’s observables,
and show that the Werner state always satisfies this condition.
Next, we use the fact that any two-qutrit state satisfying the
Collins–Gisin–Linden–Massar–Popescu inequality for arbitrary 
choices of Alice’s and Bob’s
observables admits a local hidden variable model.
\end{abstract}

\maketitle

\section{\label{sec:Intro}Introduction.}

Since the formulation of the Einstein-Podolsky-Rosen (EPR) paradox,
the question of whether quantum theory can be supplemented by 
an underlying local realistic description has motivated extensive
investigation. A decisive step in this direction was provided by J.S. 
Bell \cite{cab_Bell1964}, who showed that any local hidden
variable (LHV) model must satisfy constraints, now known as Bell 
inequalities, on observable correlations between spatially separated 
systems. 
Violations of the Bell inequality have been extensively confirmed,
see, e.g., \cite{HBDetal2015}.

Correlations violating Bell inequalities are called nonlocal.
Such correlations can be obtained from a two-partite state
$\rho_{AB}$ only when this state is entangled
\cite{BCPSW2014}. Entangled means that the state cannot be written
in the form
\begin{equation}
\rho_{AB} = \sum_{i} p_i \rho_{A}^{i}\otimes \rho_{B}^{i},
\end{equation}
where all $p_i$ are non-negative, $\sum_{i}p_i=1$
and $\rho_{A}^i$, $\rho_{B}^i$ are one-partite states.

For pure states this relation can be reversed---all pure
entangled states violate Bell inequalities
\cite{CFS1973_PureEntBell,HS1991_Bell-EPR,Gisin1991_entangled-violates-Bell}.
For mixed states the situation is more complicated.
In 1989 Werner \cite{cab_Werner1989} constructed a class of
bipartite mixed entangled states admitting a LHV model
for any local projective measurements.
The two-qudit Werner state has the following form \cite{Barret2002_LHVPOVMforWerner}
\begin{equation}
\rho_{W}^{(d\times d)} = p \frac{2}{d(d-1)} P_{\textsf{anti}}
+ (1-p) \tfrac{1}{d^2} I_d,
\label{Werner-state}
\end{equation}
where $P_{\textsf{anti}}$ is a projector on the antisymmetric
subspace of the ${\mathbb{C}}^d\otimes{\mathbb{C}}^d$ space and
$I_d$ is a $d\times d$ identity operator.
Werner proved that correlations obtained from the state
(\ref{Werner-state}) admit a LHV model for projective
measurements when $p\le\tfrac{d-1}{d}$ but the state is entangled 
for $p>\tfrac{1}{d+1}$.
Thus, the two-qubit ($d=2$) Werner state is entangled 
for $p>\tfrac{1}{3}$ while it admits a 
LHV model for $p\le\tfrac{1}{2}$.
This latter bound has been later on improved to $\approx0.696$
\cite{AGT2006_GrothendieckBell,Detal2023_GrothendieckBell,DVP2026_GrothendieckBell}.

On the other hand, for two qubits, assuming a two-measurement 
per site and two-outcome per measurement scenario, 
the Clauser-Horne-Shimony-Holt (CHSH)
\cite{cab_CHSH1969} inequality is an optimal Bell
inequality. 
A necessary and sufficient condition for CHSH violation in arbitrary
two-qubit states was derived in \cite{HHH1995_Bell-violation}.
It has been shown there that the two-qubit state $\rho^{(2\times2)}$
violates the CHSH inequality iff $u_1+u_2>1$, where 
$u_1, u_2$ are the two largest singular values of
the correlation matrix $t=[t_{nm}]$ defined as
$t_{nm}=\mathrm{Tr}[\rho^{(2\times2)} \sigma_n\otimes\sigma_m]$,
and $\sigma_i$, $i=1,2,3$ are Pauli matrices.
Therefore, the two-qubit Werner state violates the CHSH inequality
iff $p>1/\sqrt{2}\approx0.707$.

The two-qutrit Werner state we are interested in here admits
a LHV model for projective measurements when 
$p\le2/3$ \cite{BCPSW2014}.

For two-qudit systems in a two-measurement per site and $d$-outcomes 
per measurement scenario the so-called
Collins-Gisin-Linden-Massar-Popescu (CGLMP) inequality,
introduced in \cite{CGLMP_Bell_ineq_high_spin}, is optimal
\cite{Masanes2003_CGLMP-tight}.
In this framework, each party, Alice and Bob, performs one of 
two measurements $(A_1, A_2)$ or $(B_1, B_2)$, and each measurement
has $d$ possible outcomes $(0,1,2,...,d-1)$.
For $d=2$ the CGLMP inequality reduces to the CHSH inequality, 
while for 
higher dimensions it provides genuinely new constraints.
For a two-qutrit case ($d=3$) we consider here, the CGLMP
inequality takes the form
\begin{equation}\label{eq:7}
	{\mathcal{I}}_3\leq 2,
\end{equation}
where
\begin{multline}\label{eq:9}
	{\mathcal{I}}_3 =
	\Big[P(A_1 = B_1) + P(B_1 = A_2 + 1)
	+ P(A_2 = B_2) \\
	+ P(B_2 = A_1) \Big]
	-\Big [P(A_1 = B_1 - 1) + P(B_1 = A_2 )\\
	+ P(A_2 = B_2 - 1) + P(B_2 = A_1 - 1) \Big],
\end{multline}
and $P(A_a=B_b+k)$ are probabilities that the measurements $A_a$ 
and $B_b$ have outcomes that differ modulo $3$ by $k$, i.e.
\begin{equation}\label{probability-def}
	P(A_a=B_b+k) = \sum_{j=0}^{2} P(A_a=j,B_b=j+k\mod3).
\end{equation}
In \cite{Masanes2003_CGLMP-tight}, it was shown that
any two-qutrit state which does not violate the CGLMP
inequality for any choice of observables admits a LHV model
in the considered scenario.

In this paper we show that the two-qutrit Werner state satisfies
the CGLMP inequality for all values of the parameter $p$
which is equivalent to the statement that this state admits
a LHV model for projective measurements in the considered scenario.
To this end we will derive necesary and sufficient condition
for the violation of the CGLMP inequality as well as 
a very simple sufficient condition. This sufficient condition 
is expressed in terms of two-qutrit state parameters and
is easy to verify.

\section{The CGLMP inequality in a two-qutrit state}

The most general density matrix for a qutrit-qutrit system 
can be written as
\begin{multline}\label{eq:8}
\rho=\frac{1}{9}\Big[I_3\otimes I_3 + \sum_{i=1}^{8}N^A_i(\lambda_i\otimes I_3)+\sum_{i=1}^{8}N^B_j(I_3\otimes\lambda_j) \\
+\sum_{i,j=1}^{8}T_{ij}(\lambda_i\otimes \lambda_j\Big],
\end{multline}
where $\lambda_i=1,...,8$ are Gell-Mann matrices. 
The explicit form and some properties of Gell-Mann matrices are 
given in Appendix \ref{appendix.1}. 
Hermicity of these matrices implies that 
$N^A_i,N^B_j \in \mathbb{R}^8$ and
$T_{ij}=\frac{9}{4}\text{Tr}[\rho(\lambda_i \otimes\lambda_j)]$ 
is a $8\times 8$ real matrix. 
Of course $\vec{N}^A$, $\vec{N}^B$ and $T$ have to fulfill 
additional constraints imposed by semi-positive definiteness 
of the density matrix $\rho$.

Now, let us consider an arbitrary observable $M$ acting on 
a qutrit, with three distinct eigenvalues. 
In the context of Bell-type inequalities, one typically employs spin-projection observables. For spin-1 systems (i.e., qutrits), 
the eigenvalues of the spin projection operator are \( -1, 0, +1 \). Therefore, we consider an observable $ M$ with the following spectral decomposition:
\begin{equation}\label{eq:10}
	M=(-1)\Pi^M_- +(0)\Pi^M_0+(+1)\Pi^M_+.
\end{equation}
Here $\Pi^M$ are rank-1 projectors. We can now express these
projection operators in terms of observable $ M $ as follows:
\begin{equation} \label{eq:11}
	\Pi^M_{\pm}=\tfrac{1}{2}[M^2\pm M], \quad \Pi^M_{0}=I_3-M^2.
\end{equation}
One easily finds that $\text{Tr}[M]=0$ and $\text{Tr}[M^2]=2$.
The probabilities appearing in Eq. (\ref{eq:9}) have the form
\begin{equation} \label{eq:12}
	P(A=B+ k)=\text{Tr}\Big [\rho \Big(\sum_{m=-1}^{+1}\Pi^A_m \otimes \Pi^B_{m+k (\text{mod} \text{ }3)}\Big)\Big],
\end{equation}
where $\Pi_m^A$ and $\Pi^B_m$ are projectors acting on $\mathbb{C}^3$.
Now we insert Eqs.\ (\ref{eq:8},\ref{eq:11}) in 
Eq. (\ref{eq:12}) and taking into account that for $m=-1,0,1$, $\text{Tr}[\Pi_m^A] = \text{Tr}[\Pi_m^B]=1$ and 
$\sum_m\Pi^A_m = \sum_m \Pi^B_m = I_3$, we find
\begin{multline}
	\text{Tr} \Big[ 
	\sum_{m=-1}^{+1}\sum_{i=1}^{8} N_i^A 
	(\lambda_i\otimes I_3)(\Pi^A_m \otimes \Pi^B_m)
	\Big] \\
	= \text{Tr}\Big[
	\sum_{m=-1}^{+1}\sum_{j=1}^{8} N_j^B
	(I_3\otimes \lambda_j ) (\Pi^A_m \otimes \Pi^B_m)
	\Big]=0.
\end{multline}
Thus finally
\begin{multline}
	P( A_1= B_1) =  \text{Tr} 
	\Big[
	\rho\sum_{m=-1}^{+1}(\Pi^A_m \otimes \Pi^B_{m}) 
	\Big]\\
	= \text{Tr} 
	\Big\{
	\rho \big[
	\tfrac{3}{2}
	(A_1^2 \otimes B_1^2)+\tfrac{1}{2} (A_1\otimes B_1) 
	+ I_3\otimes I_3\\
	- I_3 \otimes B_1^2
	+A_1^2 \otimes I_3
	\big]
	\Big\}\\
	=  \tfrac{1}{3} + \tfrac{1}{6} \sum_{i,j=1} T_{ij}
	\Big\{
	\text{Tr} [A_1^2\lambda_i]\text{Tr}[B_1^2\lambda_j] \\
	+ \tfrac{1}{3} 
	\text{Tr}[A_1\lambda_i]\text{Tr}[B_1\lambda_j]
	\Big\}. 
	\label{probability-A1B1}
\end{multline}
In a similar way we find other
probabilities, see Eqs.
(\ref{probability-2},\ref{probability-3}). 
Inserting them in Eq. (\ref{eq:9}) 
we got the result
\begin{widetext}
\begin{multline}\label{eq:14}
{\mathcal{I}}_3 = \frac{1}{4} \sum_{i,j} T_{ij} \Big\{
\text{Tr}[A_1^2\lambda_i]
(\text{Tr}[B_1^2\lambda_j] +\text{Tr}[B_2^2\lambda_j])
+ \text{Tr}[A_2^2\lambda_i]
(\text{Tr}[B_2^2\lambda_j]-\text{Tr}[B_1^2\lambda_j])
+\frac{1}{3} \Big[
\text{Tr}[A_1\lambda_i]
(\text{Tr}[B_1\lambda_j]+\text{Tr}[B_2\lambda_j])
\\
+ \text{Tr}[A_2\lambda_i]
(\text{Tr}[B_2\lambda_j]-\text{Tr}[B_1\lambda_j])
+\text{Tr}[A_1^2\lambda_i]
(\text{Tr}[B_1\lambda_j]-\text{Tr}[B_2\lambda_j])
+ \text{Tr}[A_2^2\lambda_i]
(\text{Tr}[B_2\lambda_j]-\text{Tr}[B_1\lambda_j])
\\
+ \text{Tr}[A_2\lambda_i]
(\text{Tr}[B_2^2\lambda_j] +\text{Tr}[B_1^2\lambda_j])
+ \text{Tr}[A_1\lambda_i]
(\text{Tr}[B_2^2\lambda_j]-\text{Tr}[B_1^2\lambda_j])
\Big]\Big\}.
\end{multline}
\end{widetext}
Any $3\times 3$ observable $M$ with the spectrum ${-1,0,+1}$ 
can be obtained from $\bm {n\cdot S}$, where 
$\vec{n} \in \mathbb{R}^3$ and $S_i$ are the standard 
spin-1 matrices (for explicit form see, e.g., 
\cite{cab_Ballentine2014}),
via the unitary transformation
\begin{equation}
M=U(\vec{n}\cdot \vec{S})U^\dagger 
\quad \text{with} \quad U\in SU(3). 
\end{equation}
So we have 
\begin{equation}
\text{Tr}(M \lambda_i) = 
\text{Tr}[U(\vec{n}\cdot \vec{S})U^\dagger \lambda_i]
=\text{Tr}[(\vec{n}\cdot \vec{S})(U^\dagger\lambda_i U)],
\end{equation}
where $U\in SU(3)$.
Similarly
\begin{equation}
\text{Tr}(M^2 \lambda_i) = \text{Tr}
[(\vec{n}\cdot \vec{S})^2 (U^\dagger\lambda_i U)], 
\quad  U\in SU(3).
\end{equation}
Now we will apply the adjoint representation of group
$SU(3)$, $\gamma$, acting on $\mathbb{R}^8$
\cite{AMM1997_qutrit}. 
The matrices of this representation are defined as
\begin{equation}
	U\in SU(3) \to \gamma(U),\quad 
	U\lambda_iU^\dagger= \sum_j \gamma(U)_{ij} \lambda_j. 
\end{equation}
It holds
\begin{equation}
\gamma(U_1)\gamma(U_2) = \gamma(U_1U_2),\,\,
\gamma(U)_{ij}
=\tfrac{1}{2}\text{Tr}(\lambda_iU\lambda_jU^\dagger),
\end{equation}
and $\gamma(U) \in SO(8)$.
Thus we can write
\begin{equation}\label{eq:21}
\text{Tr}(M^2 \lambda_i) =
\text{Tr}[(\vec{n}\cdot \vec{S})^2 U^\dagger\lambda_iU]
\equiv-\tfrac{2}{\sqrt{3}}\sum_{j}\gamma(U^\dagger)_{ij}\zeta_j.
\end{equation}
and 
\begin{equation}\label{eq:22}
\text{Tr}(M \lambda_i) =
\text{Tr}[(\vec{n}\cdot \vec{S}) U^\dagger \lambda_i U]
\equiv2\sum_{j}\gamma(U^\dagger)_{ij}N_j,
\end{equation}
where we have introduced vectors 
$\vecg{\zeta}, \vec{N} \in \mathbb{R}^8$,
\begin{equation}
	\zeta_j = -\tfrac{\sqrt{3}}{2} \text{Tr}
	[(\vec{n}\cdot \vec{S})^2\lambda_j],\quad
	N_j = \tfrac{1}{2}\text{Tr}[(\vec{n}\cdot \vec{S})\lambda_j],
\end{equation}
and the factors in front of them was choosen for further convenience. 
Explicitly
\begin{multline}\label{eq:24}
\vecg{\zeta} =
\Big(
-\sqrt{\tfrac{3}{2}}n_1n_3,
-\sqrt{\tfrac{3}{2}}n_2n_3,
\tfrac{\sqrt{3}}{4}(1-3 n_3^2),
\tfrac{\sqrt{3}}{2}(n_2^2-n_1^2),\\
-\sqrt{3}n_1n_2,
\sqrt{\tfrac{3}{2}}n_1n_3,
\sqrt{\tfrac{3}{2}}n_2n_3,
\tfrac{1}{4}(3n^2_3-1)
\Big),
\end{multline}
\begin{equation}\label{eq:25}
\vec{N} =
\left(
\tfrac{n_1}{\sqrt{2}},\tfrac{n_2}{\sqrt{2}},
\tfrac{n_3}{2},
0,
0,
\tfrac{n_1}{\sqrt{2}},
\tfrac{n_2}{\sqrt{2}},
\tfrac{\sqrt{3}}{2}n_3
\right).
\end{equation}
For our reasoning it is convenient to use the standard $\star$
product defined in (\ref{star-product-def}). 
With the help of this product, we find
\begin{gather}
\label{eq:26}
\vecg{\zeta}^2=1,\quad 
\vecg{\zeta}\star \vecg{\zeta}=\vecg{\zeta},\\
\label{eq:27}
\vec{N}^2=1,\quad \vec{N} \star \vec{N} =-\vecg{\zeta},\\
\label{eq:28}
\vec{N}\cdot \vecg{\zeta} 
= -\vec{N}\cdot(\vec{N}\star\vec{N})=0,\quad
\vec{N}\star\vecg{\zeta} = -\vec{N}.
\end{gather}
Following \cite{KMSM1997}, we adopt the classification of orbits 
in the $\mathfrak{su}(3)$ Lie algebra under the adjoint action 
of the group $SU(3)$. These orbits are defined as follows:
\begin{align}  
\text{Orbit}(\vec{n}) 
= \{\vec{n}^\prime\colon \vec{n}^\prime\cdot\vecg{\lambda}
= U(\vec{n}\cdot \vecg{\lambda}) U^\dagger, U\in SU(3)\}
\label{orbit-def}
\end{align}
and are completely characterized by the invariants 
$\vec{n}^2$ and $\vec{n}\cdot(\vec{n}\star \vec{n})$. 
Excluding $n=0$, there are two classes of orbits:
(i) For three distinct eigenvalues, one obtains the generic
six-dimensional orbits $\tau(\sqrt{n^2},\theta)$ with $n^2>0$ 
and $\tfrac{\pi}{6}<\theta<\tfrac{\pi}{2}$. 
It has representative element
$\sqrt{\tfrac{\vec{ n}^2}{3}} \text{diag}
(\sin\theta+\sqrt{3}\cos\theta,
\sin\theta-\sqrt{3}\cos\theta, 
-2\sin\theta)$
and provides a realization of the coset space 
$SU(3)/(U(1)\times U(1))$.
(ii) When two eigenvalues are equal, the orbit
$\mathcal{\tau}^{(\pm)}(\sqrt{n^2})$ becomes non-generic 
and reduces to one of two four-dimensional families of coset 
space  $SU(3)/U(2)$ with the representation elements
$\sqrt{\tfrac{\vec{n}^2}{3}}(2,-1,-1)$ and 
$\sqrt{\tfrac{\vec{n}^2}{3}}(1,1,-2)$.

Now, consider sets of vectors that can be obtained 
via the adjoint action of $SU(3)$ from the vectors 
$\vecg{\zeta}$ and $\vec{N}$ given in 
Eqs. (\ref{eq:24},\ref{eq:25})
(compare Eqs. (\ref{eq:21},\ref{eq:22})):
\begin{equation}\label{eq:31}
\kappa=\{\vecg{\zeta} \in\mathbb{R}^8\colon
\vecg{\zeta}^\prime=
\gamma(U)\vecg{\zeta}, \quad U\in SU(3)\},
\end{equation}
and
\begin{equation} \label{eq:32}
\kappa^\prime =
\{\vec{N} \in \mathbb{R}^8\colon
\vec{N}^\prime = \gamma(U) \vec{N}, \quad U \in SU(3)\}.
\end{equation}
Applying the orthogonality of $\gamma(U)$, one easily sees that 
these sets form orbits of $\vecg{\zeta}$ and $\vec{N}$ 
under the adjoint action of $SU(3)$ onto its algebra
$\mathfrak{su}(3)$. Indeed
\begin{multline} \label{eq:33a}
\vec{N}^\prime\cdot \vecg{\lambda}
=\sum_{i,j} \gamma(U)_{ij} N_j\lambda_i
=\sum_{i,j} N_j \gamma(U^\dagger)_{ji}\lambda_i\\
=\sum_{j} N_j U^\dagger\lambda_j U
= U^\dagger(\vec{N}\cdot \vecg{\lambda})U,
\end{multline}
and similarly for $\vecg{\zeta}^\prime$
\begin{equation} \label{eq:33b}
\vecg{\zeta}^\prime\cdot \vecg{\lambda}
= U^\dagger(\vecg{\zeta}\cdot \vecg{\lambda})U.
\end{equation}
For the vector $\vec{N}$ given in Eq.~(\ref{eq:25}), 
we can easily find eigenvalues of the matrix
$\vec{N} \cdot\vecg\lambda$,
they are equal to: $-1,0,+1$. 
Furthermore, it can be verified that for $\theta=\tfrac{\pi}{3}$ 
the representative element of the orbit
$\tau(1,\theta)$ given below Eq. (\ref{orbit-def}) takes 
the form $\sqrt{n^2}(1,0,-1)$. 
Therefore, the set $\kappa^\prime$ is the orbit
$\tau(1,\frac{\pi}{3})$. 
For this orbit invariants have the values 
$\vec{n}^2=\vec{N}^2=1$, and $\vec{n}\cdot(\vec{n}\star\vec{n})
=\vec{N}\cdot(\vec{N}\star\vec{N})=0$ 
(c.f. Eqs. (\ref{eq:27},\ref{eq:28})). 
This orbit is a coset space of $SU(3)/(U(1)\times U(1))$. 
Thus, the action of  $SU(3)$ on this orbit is transitive.
	
In a similar manner for the vector $\vecg{\zeta}$ given in
Eq. (\ref{eq:24}), 
the eigenvalues of the matrix $\bm {\zeta\cdot\lambda}$ are:
$-\tfrac{1}{\sqrt{3}},-\tfrac{1}{\sqrt{3}},\tfrac{2}{\sqrt{3}}$.
Thus, the set $\kappa$ is the non-generic, four-parameter orbit
$\tau^+(1)$. For this orbit invariants have the values 
$\vec{n}^2= \vecg{\zeta}^2=1$, and 
$\vec{n} \cdot(\vec{n}\star\vec{n})
=\vecg{\zeta} \cdot(\vecg{\zeta}\star \vecg{\zeta})=1$ 
(c.f. Eq. (\ref{eq:26})). 
This orbit is a coset space $SU(3)/U(2)$, so the action of 
$SU(3)$ on $\tau^+(1)$ is transitive, too. 
	
Thus, the subsets $\kappa$ and $\kappa^\prime$ are invariant 
under $\gamma(U)$, and the adjoint representation acts 
transitively on both.

Now we can define pairs of vectors 
$(\vecg{\alpha}, \vec{A})$,
$(\vecg{\alpha}^\prime,\vec{A}^\prime)$,
$(\vecg{\beta},\vec{B})$,
$(\vecg{\beta}^\prime,\vec{B}^\prime)$ 
where
\begin{align} \label{eq:36}
\alpha_i 
&= \text{Tr}[A_1^2\lambda_i] 
= -\tfrac{2}{\sqrt{3}}\sum_j\gamma(U_1^\dagger)_{ij}\zeta_j,\\
A_i
&=\text{Tr}[A_1\lambda_i] 
= 2\sum_j\gamma(U_1^\dagger)_{ij}N_j,
\end{align}
with $U_1\in SU(3)$ and $\vecg{\zeta}\in \kappa$, 
$\vec{N}\in \kappa^\prime$,
and pairs 
$(\vecg{\alpha}^\prime,\vec{A}^\prime)$,
$(\vecg{\beta},\vec{B})$,
$(\vecg{\beta}^\prime,\vec{B}^\prime)$ 
are defined analogously with 
$U_2,U_3,U_4\in SU(3)$, respectively.
Note that for the pairs of vectors 
$(\vecg{\alpha}, \vec{A})$,
$(\vecg{\alpha}^\prime,\vec{A}^\prime)$,
$(\vecg{\beta},\vec{B})$,
$(\vecg{\beta}^\prime,\vec{B}^\prime)$ 
hold the same relations as those given in Eqs. (\ref{eq:26}-\ref{eq:28}) for the pair $(\vecg{\zeta},\vec{N})$.
Using these definitions 
we can rewrite Eq. (\ref{eq:14}) as
\begin{multline}
{\mathcal{I}}_3
= \tfrac{1}{3} \Big[
\vecg{\alpha} T (\vecg{\beta}+\vecg{\beta}^\prime)
+\vecg{\alpha}^\prime T (\vecg{\beta}^\prime - \vecg{\beta})
+\vec{A} T (\vec{B}+\vec{B}^\prime)\\
+\vec{A}^\prime T (\vec{B}^\prime -\vec{B})
\Big]
+\tfrac{1}{3\sqrt{3}}\Big[
\vec{A} T (\vecg{\beta} - \vecg{\beta}^\prime)
+\vec{A}^\prime T (\vecg{\beta}^\prime - \vecg{\beta})\\
+\vecg{\alpha} T ( \vec{B}^\prime -\vec{B})
+\vecg{\alpha}^\prime T (\vec{B} - \vec{B}^\prime)
\Big].
\label{I3_final}
\end{multline}

Now, our goal is to find an upper bound on the value
of ${\mathcal{I}}_3$ given in Eq. (\ref{I3_final}).
Notice that according to Eq. (\ref{eq:27}) vectors from
the pairs $(\vecg{\alpha},\vec{A})$, 
$(\vecg{\alpha}^\prime,\vec{A}^\prime)$,
$(\vecg{\beta},\vec{B})$, 
$(\vecg{\beta}^\prime,\vec{B}^\prime)$ are related with 
each other, i.e.
\begin{align}
\vecg{\alpha} & = -(\vec{A}\star\vec{A}),\quad
\vecg{\alpha}^\prime = -(\vec{A}^\prime\star\vec{A}^\prime),
\label{constaint-1}\\
\vecg{\beta} & = -(\vec{B}\star\vec{B}),\quad
\vecg{\beta}^\prime = -(\vec{B}^\prime\star\vec{B}^\prime).
\label{constaint-2}
\end{align}
Since the action of $\gamma(U)$ on the sets $\kappa$ 
and $\kappa^\prime$ is transitive, the optimization over 
all observables $A_1,A_2$ used by Alice and $B_1,B_2$ used 
by Bob is equivalent to the optimization over 
all vectors $\vecg{\alpha},\vecg{\alpha}^\prime,\vecg{\beta},
\vecg{\beta}^\prime\in\kappa$ and 
$\vec{A},\vec{A}^\prime,\vec{B},\vec{B}^\prime\in\kappa^\prime$.
with the constraints (\ref{constaint-1}-\ref{constaint-2}).
However, these constraints give a very complicated relation
between componenents of the involved vectors 
(compare (\ref{star-product-def})).
But, of course, maximum of ${\mathcal{I}}_3$ over
$\kappa$, $\kappa^\prime$ with
(\ref{constaint-1}-\ref{constaint-2})
is not grater than the free maximum of ${\mathcal{I}}_3$
over $\kappa$, $\kappa^\prime$.
Consequently, we arrive at the following theorem:
\newtheorem{theorem}{Theorem}
\begin{theorem}
If the inequality:
\begin{equation}
{\mathcal{I}}_3 \le 2,
\end{equation}
where ${\mathcal{I}}_3$ is given in (\ref{I3_final}),
holds for all vectors $\vecg{\alpha}$, $\vecg{\alpha}^\prime$, $\vecg{\beta}$, $\vecg{\beta}^\prime$,
from the set  $\kappa$ and
$\vec{A}$, $\vec{A}^\prime$, $\vec{B}$, $\vec{B}^\prime$
from the set $\kappa^\prime$ defined in 
Eqs.(\ref{eq:31},\ref{eq:32})
then
a qutrit-qutrit state (\ref{eq:8}) fulfills the CGLMP
inequality for all choices of observables used by Alice and Bob
\label{theorem-1}
\end{theorem}

\section{Optimization of ${\mathcal{I}}_3$}

To derive an upper bound on ${\mathcal{I}}_3$ from
Eq. (\ref{I3_final}) we will aply the following Lemma, proven
in \cite{CH2026_Bell2x3}:
\newtheorem{lemma}{Lemma}
\begin{lemma}\label{lemma}
Let $T$ be a real, rectangular $m\times n$ matrix. 
For arbitrary vectors $x,z\in \mathbb{R}^m$, $y,u\in \mathbb{R}^n$ 
such that $y\perp u$ it holds
\begin{equation} 
	x^T T y + z^T T u \leq \mu_1 \|x\|\|y\|+\mu_2 \|z\|\|u\|
\end{equation}\label{eq:39}
where $\mu_1,\mu_2$ are the largest singular values of the matrix $T$. 
The equality holds when $x, y$ and $z, u$ are singular vectors of $T$
corresponding to the singular values $\mu_1$ and $\mu_2$, respectively.
\end{lemma}

To apply this Lemma we will use the properties
(\ref{eq:26}-\ref{eq:28}) and group the terms in (\ref{I3_final})
into pairs containing mutually orthogonal vectors. 
Several admissible groupings of the Bell expression are possible, 
and each generally leads to a different upper bound. 
For a fixed grouping, the two largest singular values 
$\mu_1$ and $\mu_2$,
of the matrix $T$, can also be assigned to the corresponding 
terms in different ways. 
After excluding equivalent cases, the remaining bounds are compared 
under the condition $\mu_1\geq\mu_2$, and the smallest value 
is identified as the tightest bound.
We have described the technical details of this procedure
in Appendix, starting from Eq.~(\ref{grouping1}). 
As a result we obtain the following theorem:
\begin{theorem}
For the state (\ref{eq:8}) it holds
\begin{equation}
\max{\mathcal{I}_3} \le \tfrac{2}{3\sqrt{3}}
\Big[
2\sqrt{\mu_1^2+\mu_2^2+\mu_1 \mu_2}
+\sqrt{3} \sqrt{\mu_1^2+3\mu_2^2}
\Big]
\end{equation}
where ${\mathcal{I}}_3$ is given in (\ref{I3_final})
and we maximize over 
all vectors $\vecg{\alpha},\vecg{\alpha}^\prime,\vecg{\beta},
\vecg{\beta}^\prime\in\kappa$ and 
$\vec{A},\vec{A}^\prime,\vec{B},\vec{B}^\prime\in\kappa^\prime$.
\label{theorem-2}
\end{theorem}
Thus, finally, taking into account Theorems
\ref{theorem-1} and \ref{theorem-2} we arrive at the following
result
\begin{theorem}\label{theorem-3}
If for a qutrit-qutrit state $\rho$ parametrized like in 
(\ref{eq:8}) holds
\begin{equation}
2\sqrt{\mu_1^2+\mu_2^2+\mu_1 \mu_2}
+\sqrt{3} \sqrt{\mu_1^2+3\mu_2^2}
\le 3\sqrt{3}
\label{lhv-final-condition}
\end{equation}
where $\mu_1$ and $\mu_2$ are the two largest singular values 
of the matrix $T$, then the state $\rho$ is local.
\end{theorem}

\section{Locality of two-qutrit Werner state}

We obtain the two-qutrit Werner state from (\ref{Werner-state})
with $d=2$ and
$P_{\mathsf{anti}}=\tfrac{1}{2}(I_9-\textsf{SWAP})$, where
$\mathsf{SWAP}\ket{\psi}\otimes\ket{\phi}=%
\ket{\phi}\otimes\ket{\psi}$:
\begin{equation}
\rho_{W}^{(3\times3)} = \tfrac{2-p}{18}I_9 
- \tfrac{p}{6} \mathsf{SWAP},
\label{Werner-two-qutrit}
\end{equation}
and the parameter $p\in\left<-1/2,1\right>$.
The matrix $T$ (compare Eq.~(\ref{eq:8})) for the above state
reads
\begin{equation}
T(\rho_{W}^{(3\times3)}) =
-\text{diag}\,
\big[
\tfrac{3p}{4},\tfrac{3p}{4},
\tfrac{3p}{4},\tfrac{3p}{4},\\
\tfrac{3p}{4},\tfrac{3p}{4},
\tfrac{3p}{4},\tfrac{3p}{4}
\big].
\end{equation}
All singular values of this matrix are equal, therefore
we have
\begin{equation}
\mu_1(T(\rho_{W}^{(3\times3)}))=
\mu_2(T(\rho_{W}^{(3\times3)}))= \tfrac{3}{4}|p|.
\end{equation}
Inserting these values into (\ref{lhv-final-condition}),
we finnaly obtain
\begin{equation}
|p|\le 1,
\end{equation}
which is fulfilled in the whole admissible range of 
the parameter $p$.
Thus, a two-qutrit Werner state (\ref{Werner-two-qutrit})
satisfies the CGLMP inequality
in the whole region of the parameter $p$.
As was shown in \cite{Masanes2003_CGLMP-tight}, any two-qutrit 
state fulfilling the CGLMP inequality admits a LHV model for
projective measurements in the considered scenarion.
Thus, the Werner state (\ref{Werner-two-qutrit}) admits
such a model for all values of $p$.

\section{Summary and conclusions}

Concluding, we have derived a simple sufficient condition 
(Theorem \ref{theorem-3})
for a two-qutrit state to satisfy the
CGLMP inequality for arbitrary choice of Alice and Bob 
observables. 
This condition is expressed solely in terms of the
parameters of the state and maight be of interest 
in various areas of quantum theory where nonlocality 
is considered.
Next, we have shown that
the two-qutrit Werner state always fulfills this condition.
But, as proved in \cite{Masanes2003_CGLMP-tight}, 
any two-qutrit state satisfying the CGLMP inequality 
admits a LHV model in a two projective measurement per site 
and three outcomes per measurement scenario.
Therefore, we conclude that the Werner state
(\ref{Werner-two-qutrit}) admits such a model for all values of 
the parameter $p$.


\appendix

\section{Gell-Mann Matrices}\label{appendix.1}

Here we summarize the $SU(3)$ conventions employed 
throughout the paper. 
The Gell-Mann matrices $\lambda_i$ has the form
\begin{multline}
	\lambda_1=\begin{pmatrix}
		0 & 1 & 0 \\
		1 & 0 & 0\\
		0 & 0 & 0
	\end{pmatrix},\, 
	\lambda_2=\begin{pmatrix}
		0 & -i & 0 \\
		i & 0 & 0\\
		0 & 0 & 0
	\end{pmatrix},\, 
	\lambda_3=\begin{pmatrix}
		1 & 0 & 0 \\
		0 & -1 & 0\\
		0 & 0 & 0
	\end{pmatrix}, \\
	\lambda_4=\begin{pmatrix}
		0 & 0 & 1 \\
		0 & 0 & 0\\
		1 & 0 & 0
	\end{pmatrix},\,
	\lambda_5=\begin{pmatrix}
		0 & 0 & -i \\
		0 & 0 & 0\\
		i & 0 & 0
	\end{pmatrix},\,
	\lambda_6=\begin{pmatrix}
		0 & 0 & 0 \\
		0 & 0 & 1\\
		o & 1 & 0
	\end{pmatrix}, \\
	\lambda_7=\begin{pmatrix}
		0 & 0 & 0 \\
		0 & 0 & -i\\
		0 & i & 0
	\end{pmatrix},\, 
	\lambda_8=\tfrac{1}{\sqrt{3}}\begin{pmatrix}
		1 & 0 & 0 \\
		0 & 1 & 0\\
		0 & 0 & -2
	\end{pmatrix}.
\end{multline}
These matrices are Hermitian, traceless, trace-orthogonal
\begin{align}
	\text{Tr}(\lambda_i\lambda_j)=2\delta_{ij}
\end{align}
and satisfy the standard normalization convention of
$\mathfrak{su(3)}$ Lie algebra. 
Its algebraic structure is characterized by the antisymmetric 
and symmetric structure constants $f_{ijk}$ and $d_{ijk}$, 
defined through the commutation and anticommutation relations,
\begin{align}
		[\lambda_i,\lambda_j]=2i\sum_k f_{ijk}\lambda_k, \\
		{\lambda_i,\lambda_j}=\tfrac{4}{3}\delta_{ij}I_3+2\sum_k d_{ijk}\lambda_k,
\end{align}
where $f_{ijk}$ and $d_{ijk}$ have the following nonzero components:
\begin{align}
		&f_{123}=1, \quad f_{123}=f_{123}=\tfrac{\sqrt{3}}{2},\notag\\
		&f_{147}=f_{246}=f_{257}=f_{345}=f_{516}=f_{637}=\tfrac{1}{2},\\
		\text{and} \notag\\
		&d_{118}=d_{228}=d_{338}=-d_{888}=\tfrac{1}{\sqrt{3}}, \notag\\
		&d_{448}=d_{558}=d_{668}=d_{778}=-\tfrac{1}{2\sqrt{3}},\notag\\
		&d_{146}=d_{157}=-d_{247}=d_{256}=d_{344}\notag\\
		&=d_{355}=-d_{366}=-d_{377}=\tfrac{1}{2}.
\end{align}
The constant $f_{ijk}$ and $d_{ijk}$ define antisymmetric 
and symmetric bilinear operations on $\mathbb{R}^8$, respectively,
\begin{align}
(\vecg{\alpha}\wedge\vecg{\beta})_k & 
= \sum_{ij} f_{kij} \alpha_i \beta_j,\\
(\vecg{\alpha}\star\vecg{\beta})_k &
=\sqrt{3} \sum_{ij} d_{kij} \alpha_i \beta_j.
\label{star-product-def}
\end{align}
These products are important because under adjoint 
representation they transform same as vectors in $\mathbb{R}^8$,
i.e. for $\alpha,\beta\in\mathbb{R}^8$:
\begin{align}
	&(\gamma(U))\bm \alpha \wedge (\gamma(U)\bm \beta)
	= \gamma(U)(\bm \alpha\wedge \bm \beta),\\
	&(\gamma(U))\bm \alpha \star (\gamma(U)\bm \beta)
	= \gamma(U)(\bm \alpha\star \bm \beta).
\end{align}

\section{Probabilities}

All of the probabilities (\ref{probability-def}) 
can be obtained from Eq.~(\ref{probability-A1B1})
and from
\allowdisplaybreaks
\begin{multline}
P(A_2+1=B_1)
= \tfrac{1}{3} 
- \tfrac{1}{12} 
\sum_{i,j=1} T_{ij}
\Big\{
\text{Tr}[A_2^2\lambda_i] 
\text{Tr}[B_1^2\lambda_j]\\
- \text{Tr}[A_2\lambda_i] 
\text{Tr}[B_1^2\lambda_j]
+ \text{Tr}[A_2^2\lambda_i] 
\text{Tr}[B_1\lambda_j]\\
+ \tfrac{1}{3} 
\text{Tr}[A_2\lambda_i] 
\text{Tr}[B_1\lambda_j]
\Big\}
\label{probability-2}
\end{multline}
\begin{multline}
	P(A_1-1=B_2)
	= \tfrac{1}{3} 
	- \tfrac{1}{12} 
	\sum_{i,j=1} T_{ij} 
	\Big\{ 
	\text{Tr}[A_1^2\lambda_i] 
	\text{Tr}[B_2^2\lambda_j]\\
	- \text{Tr}[A_1^2\lambda_i] 
	\text{Tr}[B_2\lambda_j]
	+ \text{Tr}[A_1\lambda_i] 
	\text{Tr}[B_2^2\lambda_j]\\
	+ \tfrac{1}{3} 
	\text{Tr}[A_1\lambda_i] 
	\text{Tr}[B_2\lambda_j]
	\Big\}
	\label{probability-3}
\end{multline}
by simple substitutions.
For example $P(A_1=B_2)$ can be obtained from
(\ref{probability-A1B1}) with $B_2 \to B_1$,
$P(A_2+1=B_2)$ can be obtained from
(\ref{probability-2}) with $B_1 \to B_2$ etc.

\section{Optimization procedure}
\subsection{Grouping I}
We first consider a grouping based on the orthogonality conditions
$\vecg{\alpha}\cdot\vec{A}=0$, 
$\vecg{\alpha}^\prime\cdot\vec{A}^\prime=0$. 
Accordingly, $\mathcal{I}_3$ from (\ref{I3_final}) can be 
rearranged as follows:
\begin{multline}
3\sqrt{3}\,{\mathcal{I}}_3 = 
\sqrt{3}
\Big[
\vecg{\alpha} 
T(\vecg{\beta}+\vecg{\beta}^\prime)
+\vec{A}T(\vec{B}+\vec{B} ^\prime)
\Big]\\
+\sqrt{3} 
\Big[
\vecg{\alpha}^\prime T(\vecg{\beta}^\prime-\vecg{\beta})
+\vec{A}^\prime T(\vec{B}^\prime-\vec{B})
\Big]\\
+\Big[
\vecg{\alpha} T (\vec{B}^\prime-\vec{B})+
\vec{A} T (\vecg{\beta}-\vecg{\beta}^\prime)
\Big]\\
+\Big [
\vecg{\alpha}^\prime T (\vec{B}-\vec{B}^\prime)
+\vec{A}^\prime T(\vecg{\beta}^\prime-\vecg{\beta})
\Big].
\label{grouping1}
\end{multline}
We apply Lemma \ref{lemma} separately to each square bracket.
Now, we can assigne the two largest singular values 
$\mu_1$ and $\mu_2$ to the terms in square brackets 
in different orders, for example we have
\begin{multline}
	3\sqrt{3}\,
	{\mathcal{I}}_3 \le
	\sqrt{3}
	\Big[
	\mu_1 ||\vecg{\alpha}||\,
	||(\vecg{\beta}+\vecg{\beta}^\prime)||
	+ \mu_2 ||\vec{A}||\,
	||(\vec{B}+\vec{B} ^\prime)||
	\Big]\\
	+\sqrt{3}
	\Big[
	\mu_1 ||\vecg{\alpha}^\prime||\,
	||(\vecg{\beta}^\prime-\vecg{\beta})||
	+\mu_2 ||\vec{A}^\prime||\,
	||(\vec{B}^\prime-\vec{B})||
	\Big]\\
	+\Big[
	\mu_1 ||\vecg{\alpha}||\, 
	||(\vec{B}^\prime-\vec{B})||
	+\mu_2 ||\vec{A}||\,
	||(\vecg{\beta}-\vecg{\beta}^\prime)||
	\Big]\\
	+\Big [
	\mu_1 ||\vecg{\alpha}^\prime||\, 
	||(\vec{B}-\vec{B}^\prime)||
	+ \mu_1 ||\vec{A}^\prime||\,
	||(\vecg{\beta}^\prime-\vecg{\beta})||
	\Big]
	\label{grouping1-1}
\end{multline}
but also
\begin{multline}
	3\sqrt{3}\,{\mathcal{I}}_3 \le
	\sqrt{3}
	\Big[
	\mu_2 ||\vecg{\alpha}||\,
	||(\vecg{\beta}+\vecg{\beta}^\prime)||
	+ \mu_1 ||\vec{A}||\,
	||(\vec{B}+\vec{B} ^\prime)||
	\Big]\\
	+\sqrt{3}
	\Big[
	\mu_1 ||\vecg{\alpha}^\prime||\,
	||(\vecg{\beta}^\prime-\vecg{\beta})||
	+\mu_2 ||\vec{A}^\prime||\,
	||(\vec{B}^\prime-\vec{B})||
	\Big]\\
	+\Big[
	\mu_1 ||\vecg{\alpha}||\, 
	||(\vec{B}^\prime-\vec{B})||
	+\mu_2 ||\vec{A}||\,
	||(\vecg{\beta}-\vecg{\beta}^\prime)||
	\Big]\\
	+\Big [
	\mu_1 ||\vecg{\alpha}^\prime||\, 
	||(\vec{B}-\vec{B}^\prime)||
	+ \mu_1 ||\vec{A}^\prime||\,
	||(\vecg{\beta}^\prime-\vecg{\beta})||
	\Big].
	\label{grouping1-2}
\end{multline}
There is $2^4=16$ such assignements, leading to 16 different 
inequalities (in fact we obtain only 
6 distinct inequalities since some of them coincide).
Each of them we treat in the same manner.
First, we apply the normalization conditions 
$||\vecg{\alpha}||=||\vecg{\alpha}^\prime||=%
||\vec{A}||=||\vec{A}^\prime||=1$.
Then, we separate contributions associated with the singular 
values $\mu_1$ and $\mu_2$.
Thus, for example from Eq.~(\ref{grouping1-1}) we get,
\begin{multline}
	3\sqrt{3}\,{\mathcal{I}}_3 \le
	\mu_1
	\Big[
	\sqrt{3} ||\vecg{\beta}^\prime+\vecg{\beta}||
	+ (2+\sqrt{3}) ||\vecg{\beta}^\prime-\vecg{\beta}||
	\Big]\\
	+ \mu_2
	\Big[
	\sqrt{3} ||\vec{B}^\prime+\vec{B}||
	+ (2+\sqrt{3}) ||\vec{B}^\prime-\vec{B}||
	\Big].
	\label{grouping1-1-1}
\end{multline}
and similar inequalities for all other assignements.

Next, we want to find the maximal value of the right hand side of 
(\ref{grouping1-1-1}) (and similar inequalities obtained
for all other assignements). 
This maximal value can be found in a standard way (see, e.g.,
\cite{HHH1995_Bell-violation}).
The same method can be used for all other assignements and groupings. 
Briefly, since $||\vecg{\beta}||=||\vecg{\beta}^\prime||=1$ we have
$(\vecg{\beta}^\prime+\vecg{\beta})\perp(\vecg{\beta}^\prime-\vecg{\beta})$.
Therefore we can write
$\vecg{\beta}+\vecg{\beta}^\prime=2\cos\theta \vec{e}$,
and
$\vecg{\beta}^\prime-\vecg{\beta}=2\sin\theta \vec{e}^\prime$,
were $\vec{e}^2={\vec{e}^\prime}^2=1$, $\vec{e}\perp\vec{e}^\prime$,
$\theta\in\left<0,\pi/2\right>$.
Optimization over $\vecg{\beta},\vecg{\beta}^\prime$ 
is equivalent to optimization over $\theta,
\vec{e}\perp \vec{e}^\prime, 
\|\vec e \|=\| \vec e^\prime \|=1$.
Using this parametrization we find
\begin{equation} \label{bound-beta}
\max_{\vecg{\beta},\vecg{\beta}^\prime} 
\Big[ 
a\|\vecg{\beta}^\prime+\vecg{\beta}\|
+b\|\vecg{\beta}^\prime-\vecg{\beta}\|
\Big]
=2\sqrt{a^2+b^2}.
\end{equation}
In a similar way we get
\begin{equation} \label{bound-B}
	\max_{\vec{B},\vec{B}^\prime} 
	\Big[ 
	c\|\vec{B}^\prime+\vec{B}\|
	+d\|\vec{B}^\prime-\vec{B}\|
	\Big]
	=2\sqrt{c^2+d^2}.
\end{equation}

Using (\ref{bound-beta}-\ref{bound-B}) 
we can find maxima for all inequalities resulting from
different assignements in this grouping.
Next, all these maxima are compared under the condition
$\mu_1\ge\mu_2$ to find the tightest bound 
(i.e. we are looking for the smallest maximum).
Finally, in this grouping we arrive to
\begin{equation}
{\mathcal{I}}_3 \le 
{\mathcal{I}}_3^{\mathsf{max(I)}} 
\equiv
\frac{2\sqrt{2}\sqrt{5+2\sqrt{3}}}{3\sqrt{3}} 
(\mu_1+\mu_2).
\label{grouping-1-1-final}
\end{equation}

\subsection{Grouping II}
We rearrange $\mathcal{I}_3$ from (\ref{I3_final}) 
as follows:
\begin{multline}
	3\sqrt{3}\,{\mathcal{I}}_3 = 
	\sqrt{3}
	\Big[
	\vecg{\alpha} T(\vecg{\beta}+\vecg{\beta}^\prime)
	+\vecg{\alpha}^\prime T(\vecg{\beta}^\prime-\vecg{\beta})
		\Big]\\
	+\sqrt{3}
	\Big[
	\vec{A}T(\vec{B}+\vec{B} ^\prime)
	+\vec{A}^\prime T(\vec{B}^\prime-\vec{B})
	\Big]\\
	+\Big[
	\vec{A} T (\vecg{\beta}-\vecg{\beta}^\prime)
	+\vecg{\alpha} T (\vec{B}^\prime-\vec{B})
	\Big]\\
	+\Big [
	\vec{A}^\prime T(\vecg{\beta}^\prime-\vecg{\beta})
	+\vecg{\alpha}^\prime T (\vec{B}-\vec{B}^\prime)
	\Big].
	\label{grouping2}
\end{multline}
Next, we use the fact that normalization of vectors
$\vecg{\beta}$, $\vecg{\beta}^\prime$, $\vec{B}$, $\vec{B}^\prime$
imply the orthogonality conditions
$(\vecg{\beta}^\prime+\vecg{\beta})\perp(\vecg{\beta}^\prime-\vecg{\beta})$, $(\vec{B}^\prime+\vec{B})\perp(\vec{B}^\prime-\vec{B})$
and that $\vecg{\alpha}\perp\vec{A}$, 
$\vecg{\alpha}^\prime\perp\vec{A}^\prime$.
The same as in previous grouping, we apply Lemma \ref{lemma}
separately to each square bracket. 
Again, we can assign $\mu_1$ and $\mu_2$ to the terms in square
brackets in $2^4=16$ different ways, as a result we obtain seven 
inequivalent inequalities. For each of them
we find the maximal value of the right hand side of the
corresponding inequality and next we compare these maxima to
determine the tightest bound in this grouping.
We arrive at
\begin{multline}
	{\mathcal{I}}_3 \le 
	{\mathcal{I}}_3^{\mathsf{max(II)}}\\
	\equiv
	\frac{4\sqrt{2}}{3\sqrt{3}} 
	\sqrt{2\mu_1^2+(2+\sqrt{3})\mu_2^2+(1+\sqrt{3})\mu_1 \mu_2}.
	\label{grouping-2-final}
\end{multline}

\subsection{Grouping III}
A third inequivalent decomposition is obtained by 
the following grouping:
\begin{multline}
	3\sqrt{3}\,{\mathcal{I}}_3 =
	\Big[
	\sqrt{3} \vecg{\alpha} 
	T (\vecg{\beta}+\vecg{\beta}^\prime)
	+(\sqrt{3}\vecg{\alpha}^\prime+\vec{A}^\prime ) 
	T (\vecg{\beta}^\prime -\vecg{\beta})
	\Big]\\
	+\Big[
	\sqrt{3}\vec{A}
	T (\vec{B}+\vec{B} ^\prime)
	+(\sqrt{3} \vec{A}^\prime-\vecg{\alpha}^\prime)
	T(\vec{B}^\prime-\vec{B})
	\Big]\\
	+\Big[
	\vecg{\alpha} 
	T (\vec{B}^\prime-\vec{B})+\vec{A} 
	T(\vecg{\beta}-\vecg{\beta}^\prime )
	\Big].
	\label{grouping-3}
\end{multline}
Again we apply Lemma \ref{lemma}
separately to each square bracket.
In addition, using $\|\vecg{\alpha}^\prime\|=\|\vec{A}^\prime\|=1$
together with $\vecg{\alpha}^\prime\perp \vec{A}^\prime$, we obtain
$\left\|\sqrt{3}\,\vecg \alpha^\prime+\vec A^\prime\right\|=2$
and
$\left\|\sqrt{3}\,\vec A^\prime-\vecg \alpha^\prime\right\|=2$.
In this grouping, we have $2^3=8$ admissible singular-value
assignments (they reduce to four inequivalent bounds).
As previously, comparing these bounds under the ordering
$\mu_1\geq\mu_2$, we find the tightest bound in this grouping
\begin{multline}
	{\mathcal{I}}_3 \le 
	{\mathcal{I}}_3^{\mathsf{max(III)}}\\*
	\equiv
	\frac{2}{3\sqrt{3}} 
	\Big[
	2\sqrt{\mu_1^2+\mu_2^2+\mu_1 \mu_2}
	+\sqrt{3} \sqrt{\mu_1^2+3\mu_2^2}
	\Big].
	\label{grouping-3-final}
\end{multline}

\subsection{Tightest bound}
Finally, we compare the quantities
${\mathcal{I}}_3^{\mathsf{max(I)}}$,
${\mathcal{I}}_3^{\mathsf{max(II)}}$,
${\mathcal{I}}_3^{\mathsf{max(III)}}$ 
under the condition $\mu_1\ge\mu_2$
and find 
\begin{equation}
{\mathcal{I}}_3^{\mathsf{max(III)}} \le
{\mathcal{I}}_3^{\mathsf{max(II)}} \le
{\mathcal{I}}_3^{\mathsf{max(I)}}.
\end{equation}


%

\end{document}